\documentclass[twocolumn,aps,showpacs,preprintnumbers]{revtex4-1}

\usepackage{graphicx,latexsym,amssymb,amsmath,color,multirow,mathrsfs,epsfig,bm}
\usepackage{changes}
\usepackage{hyperref}

\hypersetup{colorlinks=true,linkcolor=blue,filecolor=magenta,urlcolor=cyan}

\begin{document}

\title{Landscape of nuclear deformation softness with spherical quasi-particle random phase approximation}

    \author{Nguyen Le Anh}
	\email{anhnl@hcmue.edu.vn}
	\affiliation{Department of Physics, Ho Chi Minh City University of Education, 280 An Duong Vuong, District 5, Ho Chi Minh City, Vietnam}
        \affiliation{Department of Theoretical Physics, Faculty of Physics and Engineering Physics, University of Science, Ho Chi Minh City, Vietnam}
	\affiliation{Vietnam National University, Ho Chi Minh City, Vietnam}
   	\author{Bui Minh Loc}
	\email{lmbui@sdsu.edu (present address)}
	\affiliation{Center for Exotic Nuclear Studies, Institute for Basic Science (IBS), Daejeon 34126, Korea}
        \affiliation{San Diego State University,
    5500 Campanile Drive, San Diego, CA 92182}
        \author{Panagiota Papakonstantinou}
	\email{ppapakon@ibs.re.kr}
	\affiliation{Rare Isotope Science Project, Institute for Basic Science, Daejeon 34047, Korea}
 	\author{Naftali Auerbach}
	\email{auerbach@tauex.tau.ac.il}
	\affiliation{School of Physics and Astronomy, Tel Aviv University, Tel Aviv 69978, Israel}

\begin{abstract}
We investigate the stability and softness of nuclei against quadrupole, octupole, and hexadecapole deformation. 
By applying the spherical Skyrme-force Hartree-Fock Bardeen-Cooper-Schrieffer quasi-particle random phase approximation, 
we diagnose ground-state deformation when imaginary solutions are obtained, i.e., the spherical ground state {\em collapses}. 
We also calculate the multipole polarizability in spherical nuclei with no collapse, as a measure of softness. 
This numerically light and theoretically sound method is found able to capture deformation patterns across the nuclide chart. 
The connection between the intrinsic shape of nuclei and the dynamics of their low-lying collective states is established and the role of shell structure is discussed.
\end{abstract}

\maketitle

\newpage
\section{Introduction}
New experimental techniques and theoretical methods continue to advance our understanding of nuclear deformation, in stable and exotic nuclei. 
Notably, the hydrodynamic model description of relativistic heavy-ion collisions is currently being used to infer deformation in nuclei.
Specifically, the elliptic flow $v_2$ and the triangular flow $v_3$, which are patterns of particle movement observed in high-energy nuclear collisions, are used in the study of quadrupole and octupole deformation, respectively \cite{starcollaboration2021, Jia2022}. 
The flow $v_4$ can also be measured, as has been done at the Relativistic Heavy Ion Collider (RHIC), and be used to study hexadecapole deformation.
 
In their ground state, most of the nuclei are not spherical but quadrupole deformed \cite{BohrMottelson1998}. A handful of octupole nuclei were confirmed in the experiment \cite{Butler2020}. Very recently, evidence of hexadecapole deformation in Uranium-238 at the RHIC was reported in Ref.~\cite{Ryssens2023}. 
Observables related to the $J^\pi = 4^+_{1}$ states of the Krypton isotopes and, in particular, on the hexadecapole degree of freedom were discussed in Ref.~\cite{SPIEKER2023}. 
One question is whether deformation values extracted through low-energy observables are consistent with the new data obtained at high energy. 
A more general question is whether we can understand in a systematic way which nuclei tend to develop hexadecapole deformation. 

A number of global theoretical models of nuclear deformation are available. 
The finite-range-droplet model (FRDM) \cite{Moller2016} is a microscopic-macroscopic model, while the fully microscopic Skyrme Hartree-Fock (HF) Bardeen-Cooper-Schrieffer (BCS) \cite{Goriely2001}, Gogny Hartree-Fock-Bogolyubov \cite{Hilaire2007} and deformed relativistic Hartree-Bogoliubov theory in continuum (DRHBc) \cite{Zhang2022} are based on a mean-field approach and take as input an effective interaction or Lagrangian. 
In mean-field approaches, the onset of deformation can be interpreted to some extent in terms of shell structure that plays an important role in the phenomenon of deformation: for example, magic nuclei are always spherical.
From the shell model, it is understood that the accumulating $p-n$ interaction strength leads to additional configuration mixing and deviations from spherical symmetry in the ground state \cite{Talmi1962, Federman1979}. The Nilsson model has been very useful for relating nuclear deformation and shell structure \cite{nilsson1955, nilsson1995book}.

From a theoretical point of view, to predict whether a nucleus is likely deformed in its ground state, 
one typically performs precise calculations of the density distributions and deformation parameters. 
In this work, we follow a different computationally light approach to examine deformation patterns throughout the nuclide chart. 
Specifically, we solve the self-consistent \textit{spherical} Skyrme HFBCS quasi-particle random phase approximation (QRPA) to obtain the excited states and response function in the quadrupole, octupole, and hexadecapole channels. 
One can then quantify the softness against deformation or diagnose static deformation by using the formal properties of the response function and the random phase approximation, respectively.

First, when no imaginary solutions are obtained, i.e., the RPA stability matrix is positive-definite (all eigenvalues are real), the static polarizability, which we might also call deformability, can be used as a measure of stiffness. 
The static polarizability is formally determined by the real part of the response function at vanishing excitation energy, which in turn is related to the inverse energy-weighted sum rule of the imaginary part of the response function via the Kramers--Kronig relation (see, e.g., \cite{Nozieres1997}).
Thus, by calculating the inverse-energy sum rule in a given channel for a spherical nucleus, we are in effect calculating the deformability of the nucleus.

Second, if an imaginary solution is obtained, that means that the spherical HFBCS state assumed to be the ground state is not actually the lowest-energy Skyrme HFBCS state \cite{Thouless1960, Thouless1961, Rowe1968, Nakada2016}:
If we were to perform HFBCS calculations with the same Skyrme functional but allowing for deformation, the ground state solution would show intrinsic deformation. 
The point here is that there is no need to perform such deformed HFBCS calculations: the fact that the (spherical) HFBCS-QRPA stability matrix is not positive-definite already indicates that the spherical HFBCS state is unstable against deformation. 
In addition, the present approach allows us to study parity-odd shapes, such as octupole, and identify regions of the nuclear chart where octupole and quadrupole deformation may coexist, perhaps leading to triaxiality. The octupole deformation was studied in this way in Ref.~\cite{LAPA23}.

We note that the excitation spectrum of deformed nuclei can be examined with formulations of QRPA which allow for deformation, for example, the self-consistent QRPA for use in axially symmetric nuclei as discussed Ref.~\cite{Terasaki2010}. Triaxially deformed nuclei are much more difficult to study microscopically, owing mainly to computational limitations.
However, we stress that the objective of the present work is not to describe the excitation spectrum but to assess static deformability. 

The manuscript is organized as follows. 
In Sec.~\ref{Sec:Method}, we describe briefly the HFBCS-QRPA formalism and how it is employed in the present work. 
In Sec.~\ref{Sec:Results}, we present and discuss our results in the octupole, quadrupole, and hexadecapole channels. 
We conclude in Sec.~\ref{Sec:Conclusion}. 

\section{Method\label{Sec:Method}}
The HFBCS-QRPA method is well known and has been long in use \cite{rowe2010book, RS80}. 
Self-consistent implementations, such as the one used here, employ the same functional in calculating the HFBCS ground state and the QRPA residual interaction. 
First, the HFBCS equations are solved for even-even atomic nuclei. The resulting HFBCS mean fields and nuclear densities are inputs to the QRPA calculation.
Then in the RPA or QRPA, the excited states $| \nu \rangle$ are the result of the operator $O^+_\nu$ acting on the correlated ground state $|\tilde{0}\rangle$ with $J^\pi = 0^+$
\begin{equation}
    | \nu \rangle = O^+_\nu |\tilde{0}\rangle.
\end{equation}

In general, the operator $O^+_\nu$ is expressed in terms of the creation and annihilation of a pair of particle-hole states ($\alpha, \beta$) that are obtained from the HF solution. The total angular momentum of the pair is $(\lambda M)$. The RPA equation \cite{RS80} is written as
\begin{eqnarray} \label{RPAmatrix}
    \sum_{\beta} \left[A_{\alpha\beta} X_\beta^{(\nu)} + B_{\alpha\beta} Y_\beta^{(\nu)} \right] &=& \omega_\nu X_\alpha^{(\nu)}, \nonumber \\
    \sum_{\beta} \left[B^*_{\alpha\beta} X_\beta^{(\nu)} + A_{\alpha\beta}^* Y_\beta^{(\nu)} \right] &=& -\omega_\nu Y_\beta^{(\nu)}.
\end{eqnarray}
The matrices $A$ and $B$ are obtained from the HFBCS single-particle state ($\alpha, \beta$) and the residual interaction which is the antisymmetrized particle-hole interaction \cite{RS80, Colo2013}.
The operator $O^+_\nu$ that excites the states $| \nu \rangle$ can be an electromagnetic field $\hat{F}(\bm r)$ with the $\lambda$-multipole component being defined by
\begin{equation}
    \hat{F}_{\lambda M}(\bm r) = e \sum_i^A r_i^\lambda Y_{\lambda M} (\hat{r}_i) \frac{1}{2} [1 - \tau_z (i)],
\end{equation}
in which the sum runs over all nucleons. $Y_{\lambda M}(\hat{r}_i)$ is the spherical harmonic function; $e$ is the proton charge; $\tau_z$ is the third component of the isospin operator.
As the ground state is $0^+$, the final total angular momentum $J$ is equal to the $\lambda$-multipole component of the operator which is $\lambda = 2, 3,$ or $4$. They are connected to the quadrupole, octupole, and hexadecapole moments of the matter distribution that are used to quantify the intrinsic shape of the nucleus. 

The degree of stability or softness of the atomic nucleus against variations is expressed via the polarizability or deformability $\mathcal{C}_\lambda$ (curvature), which is obtained from the inverse energy-weighted sum rule of the response function~\cite{RS80, Abbas1981}.
In the (Q)RPA framework, it is calculated from the $m_{-1}^\lambda$ moment
\begin{eqnarray}
    \mathcal{C}_\lambda &=& 2m_{-1}(\lambda)/A, \\
    m_{-1}(\lambda) &=& \sum_n |\langle \lambda_n || \hat{F}_\lambda|| \tilde{0} \rangle |^2 E_n^{-1}. \label{curvature}
\end{eqnarray}
The larger $\mathcal{C}_\lambda$ (W.u./MeV) is, the less stiff the system is against deformations. For example, in the case of octupole deformation $\mathcal{C}_3$ is around $1$ W.u./MeV for $^{96}$Ru, which is not considered especially soft. In contrast, it is $4$ W.u./MeV (even larger) for $^{96}$Zr which is an example of octupole softness nuclei \cite{LAPA23}.

Our method in the study of nuclear deformation softness is described as follows.
Solving the QRPA equation (\ref{RPAmatrix}) gives us the energies $\omega_\nu$ and the wave functions ($X^{(\nu)}, Y^{(\nu)}$) of the excited states, where $\nu$ enumerates those states.
When the ground state is stable against the particle-hole perturbations, the QRPA equation (\ref{RPAmatrix}) has only real solutions $\pm \omega_\nu$~\cite{Thouless1960, Thouless1961}.
If, however, a deformed HFBCS state is energetically preferable to the spherical one, the QRPA solutions in the relevant multipolarity will include imaginary values. 
We call the occurrence of such an instability a ``collapse"  following Ref.~\cite{Abbas1981}, where the octupole-deformed softness of $^{96}$Zr was first recognized. 

In Ref.~\cite{LAPA23} we presented results of spherical QRPA calculations in the octupole channel using various functionals. 
For certain nuclei, we found that the results showed special sensitivity to the functional used, namely the values for $\mathcal{C}_3$ varied greatly, with the low-lying collective state at very low energy and in some cases ``collapsed". 
Such results indicate that the nucleus is likely soft or even unstable against specific deformation. 
$^{96}$Zr is an example of a \textit{soft nucleus}, while $^{96}$Ru is an example of a \textit{stiff nucleus} against octupole deformation. Doubly magic nuclei such as $^{208}$Pb are examples of stiff nuclei to all kinds of deformation. 
In the next section, we highlight soft as well as stiff nuclei in the octupole, quadrupole, and hexadecapole channels. 

We applied the self-consistent HFBCS-QRPA calculation using the computer program presented in Refs.~\cite{Colo2013, Colo2021}. Skyrme forces SLy5 \cite{Chabanat1998} and SkM$^*$ \cite{Bartel1982} are used. We computed the energies of first excited states $E(2^+_1), E(3^-_1)$, and $E(4^+_1)$ for all even-even nuclei with experimental data available in the NuDat database \cite{NuDat}. All used parameters are described in detail in Ref.~\cite{Colo2013, Colo2021}. When the spherical calculation for a given atomic nucleus has an imaginary solution, we conclude that the spherical ground state is predicted unstable against the QRPA operator (quadrupole, octupole, or hexadecapole), in other words, the nucleus is predicted static deformed. In the case of the quadrupole operator, as expected, we find the majority of nuclei to be deformed. 
When all QRPA solutions are real, we use the multipole polarizability as a comparative measure of softness. 
Softness against deformation can be concluded also when the first excited state is found at real but very low energy and highly collective. 
When, for example, the transition strength in W.u. exceeds half the mass number, i.e., enhancement), we can quite confidently classify the nucleus as soft.

\section{Results and Discussion\label{Sec:Results}}
\subsection{Octupole deformation softness}
\begin{figure}[t]
    \centering
    \includegraphics[width=0.5\textwidth]{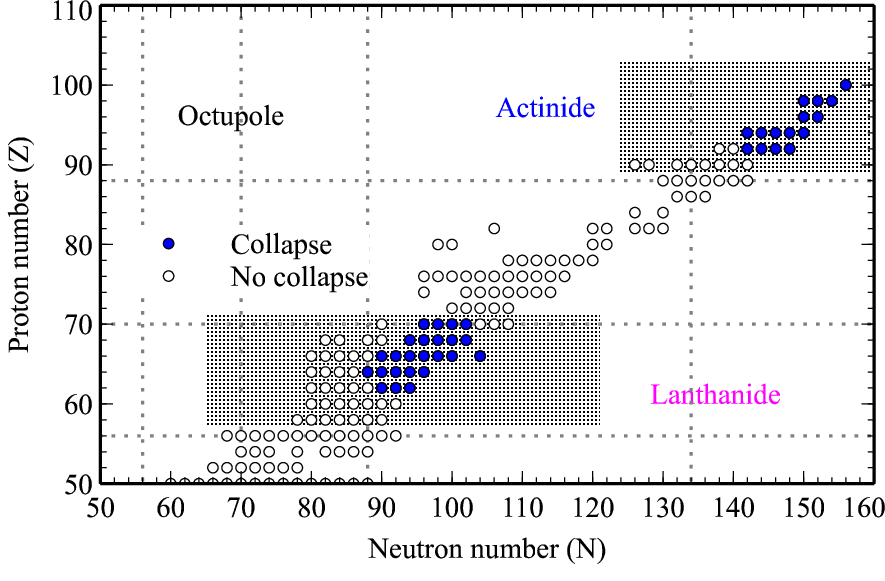}
    \caption{Octupole deformation landscape with the SLy5 force according to the presence (collapse) or not (no collapse) of imaginary solutions. Lanthanide and actinide regions are shaded. The dotted lines indicate the octupole-driving numbers. This figure has been essentially adopted from Ref.~\cite{LAPA23}.
  \label{fig:LandscapeE3}
}
\end{figure}
We begin with the octupole deformation, which is of negative parity and much less studied than even-parity shapes. 
Generally, negative-parity nuclear shapes are of interest in searches for physics beyond the standard model of particle physics ~\cite{Auerbach1996, Spevak1997} (see \cite{Pospelov2005,Butler2020} for reviews).
Motivated by the results of recent heavy-ion experiments, which indicate octupole softness in $^{96}$Zr~\cite{starcollaboration2021,Zhang2022b},  
we examined potential static octupole deformation softness across the nuclear chart in Ref.~\cite{LAPA23}. 

The results were consistent with the concept of octupole magic numbers: 
The development of $3^-$ octupole deformation softness requires the presence of a particle and a hole state of opposite parity but in close proximity and with angular momenta such that they can couple to total angular momentum number $3$. 
Such particle-hole pairs are very specific and owe their existence to the spin-orbit splitting~\cite{LAPA23,Abbas1981}. 
For example, octupole deformation areas are found in the lanthanide and actinide regions where the $1h_{11/2}$ or $1i_{13/2}$ hole orbitals come energetically close to opposite-parity particle states $2d_{5/2}$ and $2f_{7/2}$, respectively. 

In Figure \ref{fig:LandscapeE3} that essentially adopted from Ref.~\cite{LAPA23}, we recall the resulting landscape for octupole deformation softness. 
The calculation was done only for nuclei for which experimental data are available for the energy and strength of the $3^-_1$ transition to the ground state. 
The blue circles in Figure \ref{fig:LandscapeE3} indicate nuclei where we find ``collapse" in the SLy5 spherical QRPA calculation, i.e., the presence of an imaginary solution, indicating instability of the spherical shape against octupole deformation in this model. More generally, it indicates softness against octupole deformation~\cite{LAPA23}. 

\begin{figure}[t]
    \centering
    \includegraphics[width=0.5\textwidth]{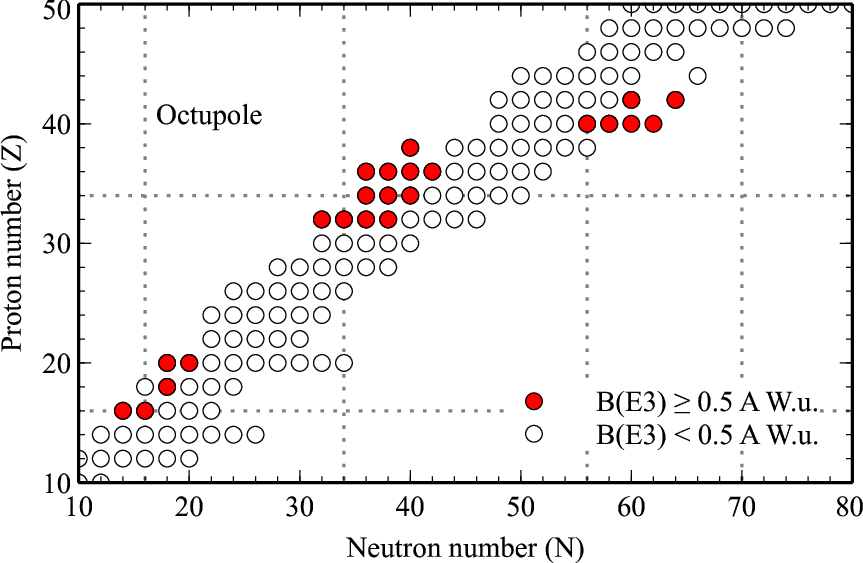}
        \caption{Light nuclei with strong enhanced octupole transition strength, $B(E3)> 0.5~A$~W.u., calculated with the SLy5 force. The dotted lines indicate the octupole-driving numbers.
    \label{fig:BE3A}}
\end{figure}
Also in Ref.~\cite{LAPA23}, it was pointed out that no ``collapse" was obtained in light nuclei. However, there were cases where enhanced $B(E3)$ transition strength was found at low energy and correspondingly high polarizability. 
A high polarizability indicates strong octupole correlations and a degree of softness.  
According to Ref.~\cite{LAPA23}, nuclei with enhanced transition strength are highlighted in Figure~\ref{fig:BE3A}. Not only the ``collapse'' but also the enhancement of the transition strength are signatures of nuclear deformation softness.

In addition, the transition from octupole deformation to octupole vibrations has been suggested within two different models (see Refs.~\cite{bizzeti2004,bizzeti2008} and \cite{bonatsos2005,lenis2006}), with $^{226}$Ra and $^{226}$Th near the actinide region appearing to be close to the transition point. The experimental identification of stable octupole deformation in $^{224}$Ra \cite{gaffney2013} supports these predictions.

In the following, we analyze the quadrupole and hexadecapole deformation analogously.

\subsection{Quadrupole deformation softness}
Next, we demonstrate how the present approach works in the case of the well-established quadrupole deformation. 
Quadrupole deformation is very common throughout the nuclear chart and has been the most extensively studied shape~\cite{Moller2016,Zhang2022}. 
Experimental data for the first excited quadrupole state, $2^+_1$, have also been systematically documented~\cite{Pritychenko2014}. 
Its properties are closely related to shell structure, for example, a high energy is associated with magicity. 
Systematic QRPA calculations using SLy4 and SkM$^*$ force for the $2^+_1$ state in spherical nuclei were carefully discussed in Ref.~\cite{Terasaki2008}. 
In that work, $155$ (SLy4) and $178$ (SkM$^*$) spherical nuclei were selected and studied. 

As already explained, in this work we solve the spherical QRPA equations for all nuclei for which relevant data exist, regardless of sphericity: 
If no imaginary solution is obtained, which means that the model does not predict quadrupole deformation, the energy and strength of the $2^+_1$ state and the quadrupole polarizability can be used as measures of the nucleus's softness against quadrupole deformation; if we obtain an imaginary solution, then the nucleus is predicted to be deformed in its ground state. 

\begin{figure}[t]
    \centering
    \includegraphics[width=0.5\textwidth]{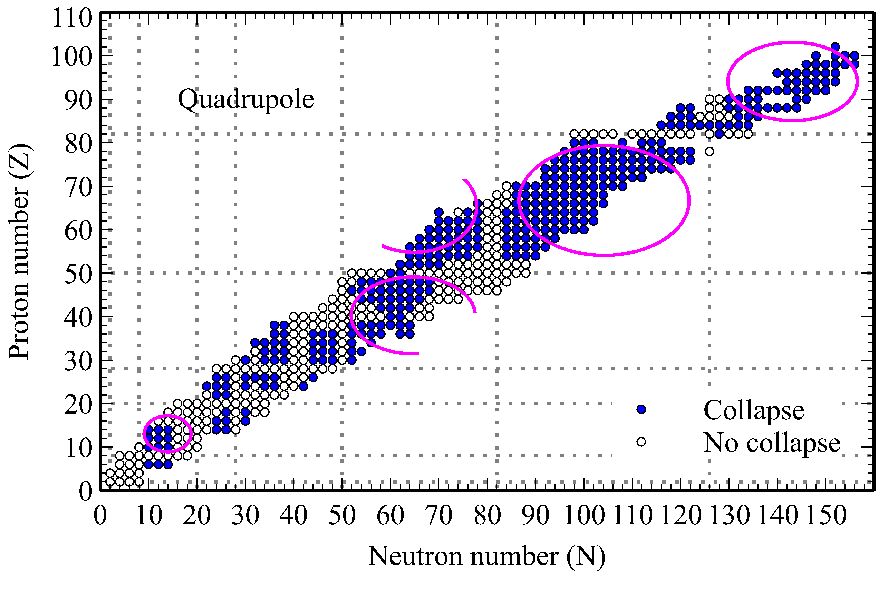}
    \caption{Quadrupole deformation landscape with the SLy5 force according to the presence (collapse) or not (no collapse) of imaginary solutions. The mid-shell deformed regions are highlighted by circle or circular segments. The magic numbers are shown in dotted lines.
   \label{fig:LandscapeE2_midshell}
}
\end{figure}

The difference from the octupole case is that because of the positive parity and low angular momentum of the $2^+$ state, many particle-hole components are available for forming a $2^+$ quadrupole state
and many particle-hole pairs may lie close energetically in open-shell nuclei leading to softness or to the instability of the spherical shape. 
As a result, the shell-structure origin of quadrupole deformation is not as clearly identified as in the case of octupole deformation. 
It is also unsurprising that quadrupole deformation is so common across the nuclear chart. 
Figure \ref{fig:LandscapeE2_midshell} shows the quadrupole deformation landscape with the mid-shell deformed regions highlighted by circle or circular segments. 
The most strongly deformed nuclei that are well-known in the literature, namely  
 $^{20-24}$Ne, $^{22-26}$Mg, $^{26,28}$Si, $^{98-102}$Sr, $^{102}$Zr, and $^{152}$Ce, can be identified in Figure \ref{fig:LandscapeE2_midshell}, too, as deformed.

Regarding regions in which quadrupole deformation appears, it would be interesting to mention the parameter-independent predictions of the $N_pN_n$ scheme \cite{casten1985} and the so-called $P$-factor \cite{casten1987}, which is determined from the number of valence protons and valence neutrons counted from the nearest closed shell. Nuclei with $P>5$ are expected to exhibit quadrupole deformation (see, for example, Figure 21 in Ref.~\cite{casten2007}).
 
In conclusion, our approach based on self-consistent QRPA captures the same region of quadrupole deformation as well-known in literature such as in Ref.~\cite{Moller2016} using the microscopic-macroscopic model.

\subsection{Hexadecapole deformation softness}
\begin{figure}[t]
    \centering
    \includegraphics[width=0.5\textwidth]{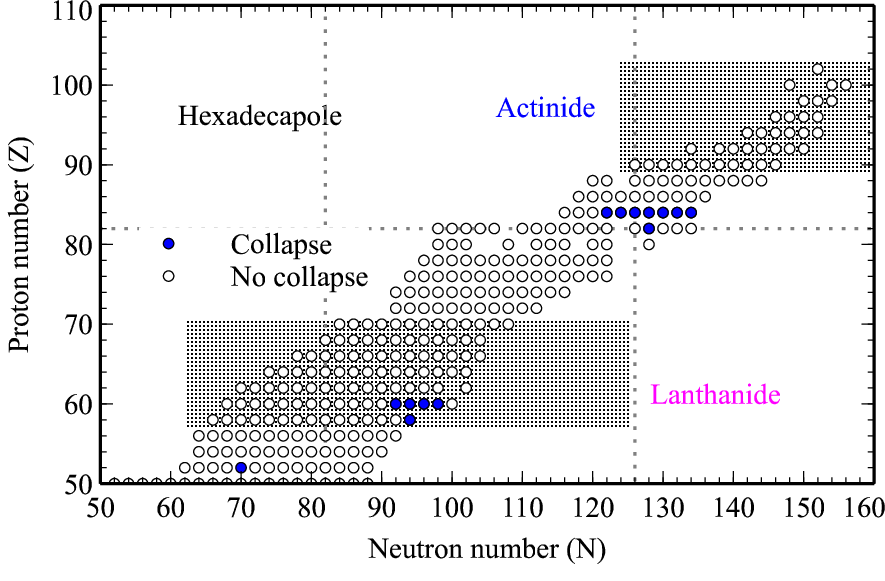}
    \caption{Hexadecapole deformation landscape with the SLy5 force according to the presence (collapse) or not (no collapse) of imaginary solutions. The dotted lines indicate the magic numbers. Lanthanide and actinide regions are shaded.
    \label{fig:LandscapeE4}
  }
\end{figure}

Hexadecapole deformation is not as thoroughly discussed in the literature as quadrupole deformation. 
Although the parity in both cases is positive, the number of single-particle pairs that can form a $J = 4$ state is more limited. 
In addition, most of those pairs can couple to $2^+$ states as well, so that hexadecapole softness or deformation is expected to be favored in nuclei with strong quadrupole softness or deformation. 
Figure~\ref{fig:LandscapeE4} shows that it is around neodymium ($Z = 60$) and polonium ($Z = 84$) that QRPA collapses in the $4^+$ channel.
Observing the single-particle shell structure, the hexadecapole-soft nuclei appear after closing the major shells. Regions of softness, however, are more extended.

\begin{figure}[b]
    \centering
    \includegraphics[width=0.5\textwidth]{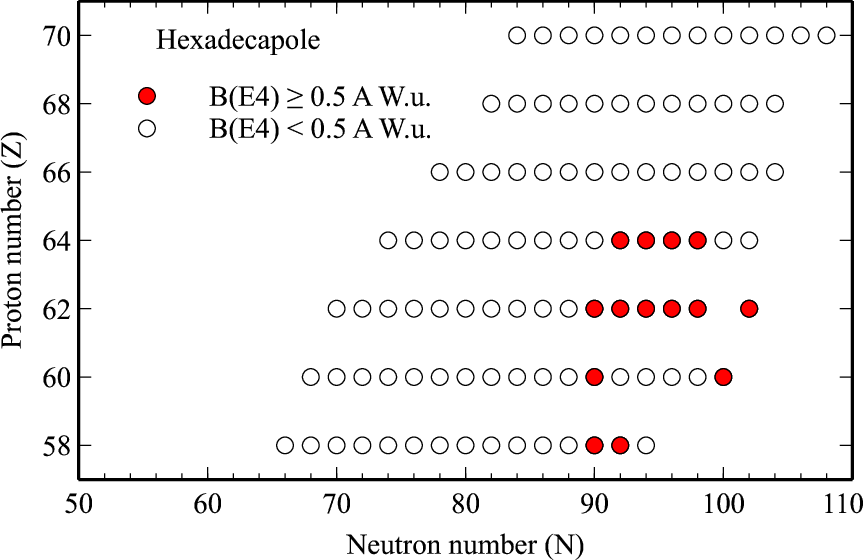}
    \caption{Nuclei with especially enhanced hexadecapole transition strength, $B(E3)> 0.5~A$~W.u. in the lanthanide region ($Z=57-71$), as calculated with SLy5 HFBCS-QRPA. These are certain Ce, Nd, Sm, and Gd isotopes with $N\geq 90$. 
 \label{fig:BE4A}
  } 
\end{figure}

The area where there are enhancements of $B(E4)$ transition strength is shown in Figure~\ref{fig:BE4A}. 
The extension of this area to the neutron-rich nuclei can provide us with more information on the single-particle structure of the whole region.
The $Z, N=59-70, 96-112$ regions of the nuclear chart are interesting also in the context of shape coexistence, as they are candidates for the dual-shell mechanism proposed in Ref.~\cite{Martinou2021}.

\begin{figure}[t]
    \centering
    \includegraphics[width=0.5\textwidth]{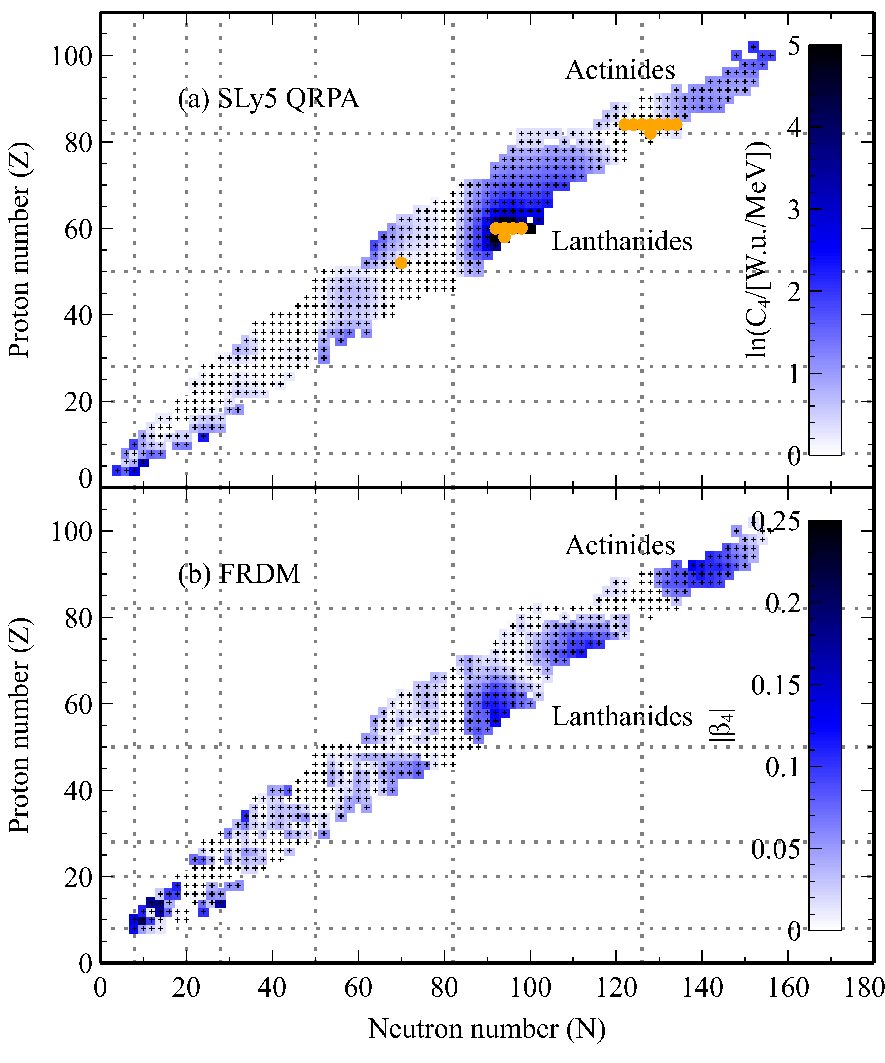}
    \caption{(a) Landscape of the polarizability $\mathcal{C}_\lambda$ with the SLy5 force. The yellow color indicates nuclei for which we have not obtained a result for the polarizability because static deformation is predicted (collapse of spherical QRPA). (b) The absolute value of the hexadecapole deformation parameter as predicted by the FRDM model~\cite{Moller2016} for the nuclei studied in this work. 
    \label{fig:E4-FRDM}}
\end{figure}

In Figure~\ref{fig:E4-FRDM} we show a broader range of $\mathcal{C}_4$ and compare visually the resulting landscape with predictions for the absolute value of the hexadecapole deformation parameter based on the FRDM model~\cite{Moller2016}. 
The regions of deformation or softness correspond almost one-to-one to our results. Regarding the division of the Lanthanide region into two islands instead of one, we note that one of the islands corresponds to positive and one to negative values of $\beta_4$, a distinction that we cannot make with our approach.

\subsection{Nuclear deformation landscape}
\begin{figure}
    \centering
    \includegraphics[width=0.5\textwidth]{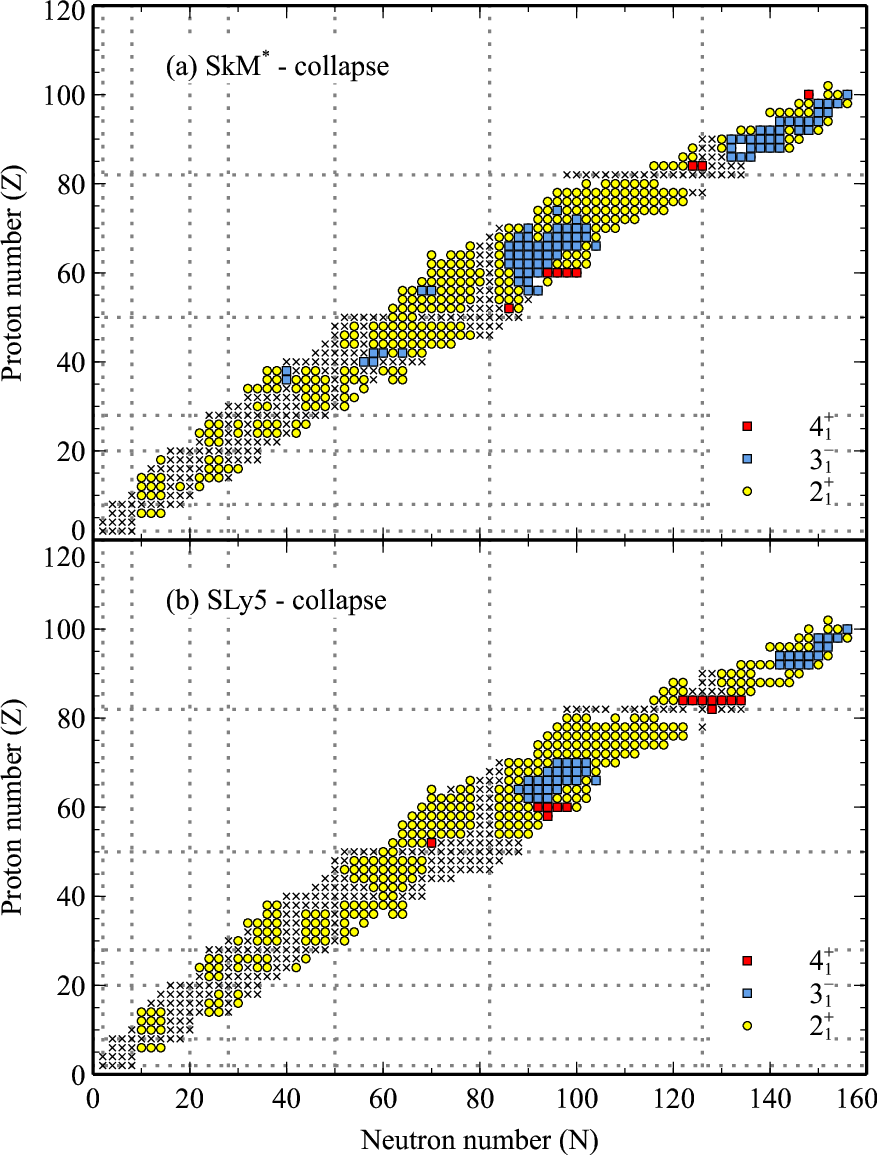}
    \caption{Nuclear deformation landscape with different SkM$^*$ (a) and SLy5 (b) forces. The collapses are in color. The cross symbol presents nuclei without any ``collapse" in the completed diagnostic. The dotted lines indicate the magic numbers.}
    \label{fig:CollapseE234}
\end{figure}

Our approach is applicable to the deformation softness and stiffness of any multipolarity in atomic nuclei including quadrupole, octupole, and hexadecapole modes from light to heavy nuclei, as we have seen.
Figure \ref{fig:CollapseE234} shows the overview result of the ``collapse'' throughout the nuclide chart with two different Skyrme forces, SkM$^*$ in Figure~\ref{fig:CollapseE234}(a) and SLy5 in Figure~\ref{fig:CollapseE234}(b). One can see that the octupole or hexadecapole ``collapse'' coincides with the quadrupole ``collapse''. This means that the octupole and the hexadecapole deformation are always on top of the quadrupole-deformation background. As mentioned in Ref.~\cite{Terasaki2008}, it was surprising to find that the two Skyrme forces disagreed significantly on which nuclei are spherical. However, looking at the landscape in Figure~\ref{fig:CollapseE234}, we observe that the predicted areas of deformation are similar, even though they are not identical. 
The important conclusion is that changes predicted by different models in the single-particle shell structure may differ somewhat, leading to differences in the deformation landscape as predicted with different Skyrme forces \cite{Terasaki2008} and thus reaffirming the role of the shell-driving mechanism in the evolution of nuclear deformation.
 
Note that the original magic numbers, $2, 8, 20, 28, 50, 82$, and $126$ \cite{Mayer1948, Haxel1949} enhance stability and make nuclei spherical and stiff against nuclear deformations. In the context of nuclear deformations, the original magic numbers are interpreted as the stiff-multipole numbers.
The presence of strong multiple correlations decreases the stiffness of the atomic nucleus, making them soft to the corresponding deformation. Therefore, the octupole-driving particle numbers or octupole-magic numbers ($16, 34, 56, 88$, and $134$) represent soft-octupole numbers. The evolution of octupole magic numbers in neutron-rich nuclei could be an interesting subject for future work. 

\begin{figure}
    \centering
    \includegraphics[width=0.5\textwidth]{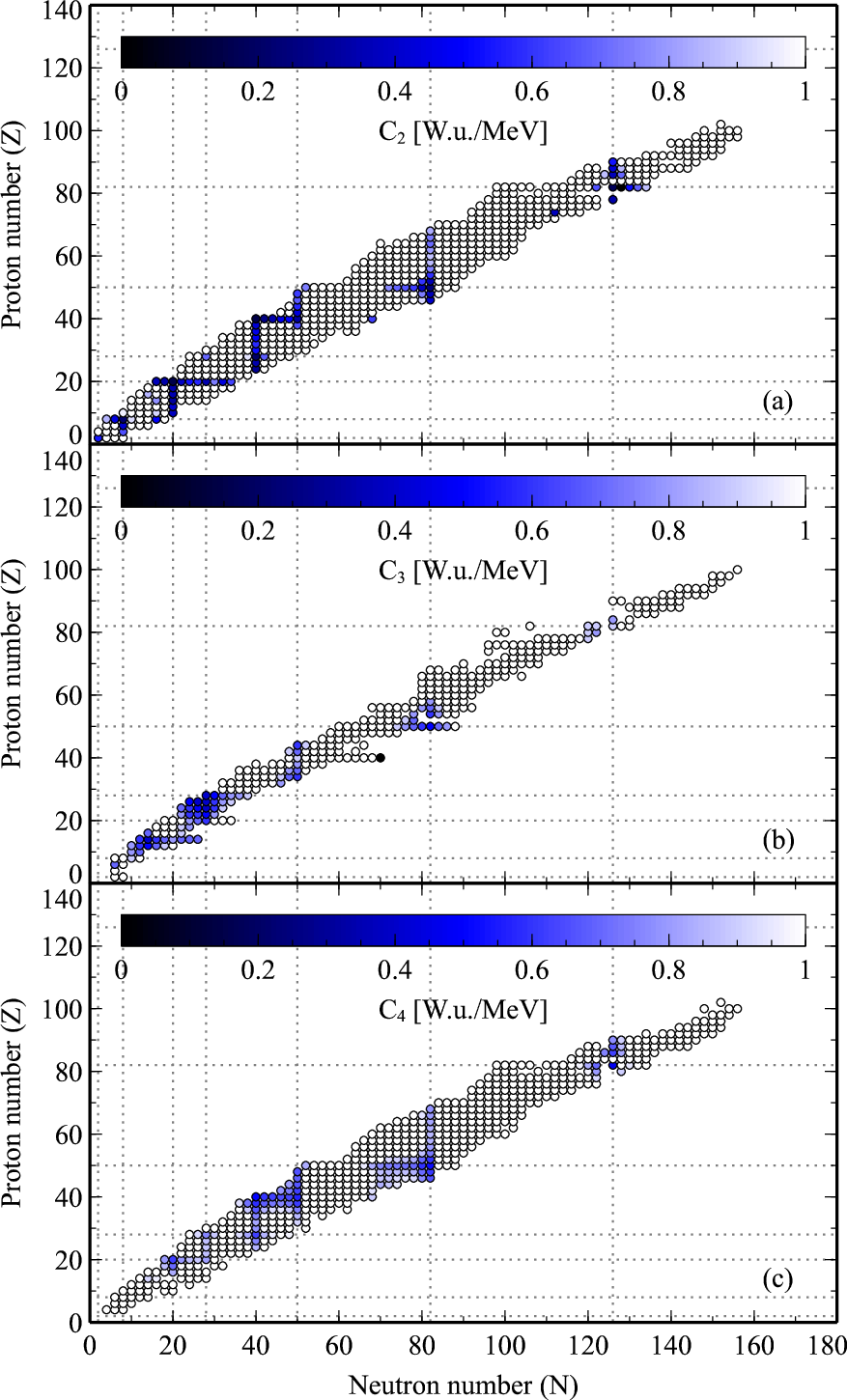}
    \caption{Landscape for nuclear \textit{stiffness} against multipole deformation through the polarizability $\mathcal{C}_\lambda$ with the SLy5 force.
    \label{fig:alpha_stiff}}
\end{figure}

While the stiff-quadrupole-deformed nuclei corresponding to the original magic number are trivial (Figure~\ref{fig:alpha_stiff}(a)), the stiff-octupole and the stiff-hexadecapole magic numbers are less recognized and paid attention to so far. In Figures~\ref{fig:alpha_stiff}(b) and (c), therefore, we show also the value of $\mathcal{C}_{\lambda = 3,4}$ (Eq.~\eqref{curvature}) throughout the nuclide chart. The quadrupole, octupole, and hexadecapole magic nuclei are in the colored regions.

As octupole and hexadecapole softness coincide with quadrupole softness, different types of magic numbers compete. As an example, in the case of $^{84}_{34}$Se$_{50}$, our result shows that the effect of octupole deformation softness is surpassed by the stiffness of the closed shell structure. 
Our method based on the fully self-consistent HFBCS-QRPA can diagnose deformation softness against one multipolarity at a time. 
For example, we cannot conclude whether a specific nucleus will show, at the same time, quadrupole and octupole deformation or quadrupole and hexadecapole. 
However, in qualitative terms at least, our approach can reliably diagnose softness in atomic nuclei and deformation patterns in the nuclide chart. 
Our results can serve as a guide to more quantitative explorations.

In soft nuclei, low-lying states exhibit greater collectivity, making them more effectively described within the RPA framework. The low-lying collective energy-level spectra below a few MeV reveal the shape of the nuclear ground state.
Our analysis with the Skyrme HFBCS-QRPA emphasizes the role of shell driving mechanism in the evolution of nuclear deformation. 
It is reminded that the spin-orbit interaction plays a crucial role in determining the single-particle shell structure, magic numbers, and therefore nuclear deformations.
The strength of the spin-orbit interaction is determined empirically and is not fully understood quantitatively, especially for weakly bound nuclei \cite{Kay2017}.

From the shell-model perspective, it has been shown that interactions between neutrons and protons are primarily responsible for causing nuclear deformation and that the same physical explanation holds across the nuclear chart~\cite{Talmi1962, Federman1979}. 
The review of the microscopic origin of nuclear deformation from different approaches was discussed in Ref.~\cite{Nazarewicz1994}. 
The spin-orbit-like shells drive also the mechanism for shape coexistence discussed in Ref.~\cite{Martinou2021}. 
In general, the degeneracy of eigenvalues of a single-particle Hamiltonian around the Fermi level leads to instability concerning shape vibrations. 

For completeness, in the Supplementary Material, we show the full range of $\mathcal{C}_{\lambda}$ values and indicate nuclei with collapse for all three multipolarities, as calculated with the SLy5 functional. We also show the overview of the nuclear deformation softness landscape with three different Skyrme forces based on the energy of the lowest excited state in each channel. Finally, we provide tables with the numerical results obtained and used in the figures.

\section{Conclusion and future perspectives \label{Sec:Conclusion}}
The \textit{spherical} QRPA is a numerically light and theoretically sound method for diagnosing potential softness or deformation in nuclei, especially octupole, hexadecapole, and higher-multipolarity shapes, which are much less studied than quadrupole shapes and typically require special microscopic approaches. 
It can capture the patterns of deformation in the nuclide chart and can help identify easily areas of potential interest. 
Candidates for static quadrupole-octupole deformed nuclei, as judged from the simultaneous softness or collapse in both channels, were reported in Ref.~\cite{LAPA23}. 
In this work, $^{122}$Te, $^{152}$Ce, $^{152-158}$Nd, $^{210}$Pb, and $^{206-218}$Po are found to be possible candidates for static quadrupole-hexadecapole deformation. 

Nuclear deformation is an ongoing research. Nowadays, new exotic beam facilities have been built around the world.
The impact of the nuclear shape on the drip lines was recently discussed in Ref.~\cite{tsunoda2020}. Theoretically, the determination of the drip lines has been updated up to heavy nuclei from nuclear density functional theory \cite{neufcourt2020} and up to Fe
through the valence-space formulation of the in-medium similarity renormalization group \cite{stroberg2021}. Moreover, the difficulty in assigning meaningful effective values to the shape parameters beta and gamma because of large fluctuations, questioning the spherical shape of doubly magic nuclei has been recently raised \cite{poves2020}. Our work can serve as a starting point for more detailed investigations in the future.

\section*{Acknowledgements} 
N. L. A. was funded by the Master, PhD Scholarship Programme of Vingroup Innovation Foundation (VINIF), code VINIF.2023.TS.003. B. M. L. was supported by the Institute for Basic Science (IBS-R031-D1) and by SSAA under DOE NNSA Contract No. DE-NA0004075. P. P. was supported by the Rare Isotope Science Project of the Institute for Basic Science funded by the Ministry of Science, ICT and Future Planning and the National Research Foundation (NRF) of Korea (2013M7A1A1075764).

\bibliography{refs.bib}

\newpage

\newpage

\end{document}